\documentclass[prd,aps,floatfix,nofootinbib,preprint ,tightenlines,superscriptaddress]{revtex4}
\usepackage{dcolumn}
\usepackage{amsmath}
\usepackage{latexsym}
\usepackage{graphicx}
\usepackage{bm}

\def\svev#1{\left\langle #1\right\rangle}       

\def\Tr{{\rm Tr}\,}

\newcommand{\bee}{\begin{equation}}
\newcommand{\ee}{\end{equation}}
\newcommand{\beea}{\begin{eqnarray}}
\newcommand{\eea}{\end{eqnarray}}

\begin{document}

\title{Topological susceptibility in QCD with two flavors and 3-5 colors -- a pilot study}

\author{Thomas DeGrand}
\affiliation{
Department of Physics, University of Colorado,
        Boulder, CO 80309 USA}
\email{thomas.degrand@colorado.edu}

\date{\today}

\begin{abstract}
I present a calculation of the topological susceptibility $\chi_T$
 in $SU(N_c)$
gauge theory with $N_c=3-5$ colors and $N_f=2$ degenerate flavors of fermions.
The results lie on a common curve when expressed in terms of the
combination $N_c m_{PS}^2 t_0$ where $m_{PS}$ is the pseudoscalar meson mass
and $t_0$ is the flow parameter.
 $\chi_T$ approaches its quenched value as the pseudoscalar mass becomes large.
 The lattice simulations use clover fermions. They are
 done at a single lattice spacing, roughly matched across $N_c$, and over a restricted
range of fermion masses.
\end{abstract}
\maketitle

\section{Introduction, motivation, and results}
Real world QCD, with $N_c=3$ colors, shares many features with its $N_c\rightarrow \infty$ limit.
Large $N_c$ expectations \cite{tHooft:1973alw,tHooft:1974pnl} mostly arise from graph counting,
in that only planar diagrams survive in the  large $N_c$ limit.
The consequences of these expectations are often applied to nonperturbative observables, like
masses or matrix elements, and these applications are qualitatively satisfied by experimental data.

Nonperturbative predictions really need nonperturbative checks, and there is a small lattice literature
of simulations of QCD with $N_c > 3$.   (See \cite{Lucini:2012gg,DeGrand:2012hd,Bali:2013kia,DeGrand:2016pur,GarciaPerez:2020gnf}
for a selection of reviews and original work.)  Simulation results generally agree with expectations.
This note is another check of large $N_c$ counting.
It is a calculation of the topological susceptibility
with two flavors ($N_f=2$) of degenerate mass fundamental representation fermions and $N_c=3$, 4, and 5 colors.

There are at least three approaches in the literature for studying
large $N_c$ QCD with lattice methods. The largest $N_c$ values are attained by
 assuming volume independence
and simulating on small spatial volumes (see  \cite{GarciaPerez:2020gnf} for a recent review).
Fairly large $N_c$ values (up to $N_c=17$)
have been reached
doing quenched simulations  \cite{Bali:2013kia} on large volumes. Simulations using full QCD, with dynamical fermions, in large volumes
are much more costly. However, part of the large $N_c$
phenomenology is that fermion effects die away at large $N_c$. To see their effects die away,
 it is necessary to include dynamical fermions from the start.

Naive large $N_c$ counting does not address possible effects depending on the fermion mass $m_q$. For many processes,
$N_c$ and $m_q$ effects approximately factorize: $Q(N_c,m_q) \sim N_c^p f(m_q)$.
Examples include meson masses ($m_H$ vs $m_q$), decay constants, and even baryon masses
($M_B(N_c,m_q,J)= N_c m_0(m_q) + (J(J+1)/N_c) B(m_q) +\dots$ for angular momentum $J$). The qualitative agreement
 of full
QCD ($N_c=3$ with dynamical fermions) and quenched QCD (replacing a dynamical fermion by a quenched valence one)
is a consequence of this factorization.

But there are (at least) two cases where this factorization should not occur. These cases occur
for chiral observables and follow from the scaling of
the pseudoscalar decay constant $f_{PS}$ and condensate $\Sigma$: $f_{PS} \propto N_c^{1/2}$ and
 $\Sigma \propto N_c$.
(The behavior of $f_{PS}$ is directly tested on the lattice; the second relation is only known
indirectly:  the squared pseudoscalar mass divided by the fermion mass $m_{PS}^2/m_q$ is 
seen to be independent
of $N_c$ and this ratio is also proportional to $\Sigma/f_{PS}^2$.)
The first case involves quantities scaling as $m_{PS}^2/f_{PS}^2 \propto m_{PS}^2/N_c$ or $m_q/N_c$.
Examples include higher order corrections to chiral observables,
$O=O_0(1 + C (m_{PS}^2/f_{PS}^2) \log(m_{PS}^2/\Lambda^2)+\dots )$.
These are typically hard to see in simulations because they are sub-leading corrections.
One example, though, has been reported in  Ref.~\cite{DeGrand:2017gbi},
 the dependence of the gradient flow scale $t_0$
on $m_{PS}^2/N_c$ as described by Golterman and Shamir \cite{Bar:2013ora}.

The second case is the subject of this note: the topological susceptibility $\chi_T$.
It has very different behavior  in the quenched limit and at small fermion mass.
In the former case  $\chi_T$ is a constant (call it $\chi_Q$), which is nearly independent of $N_c$.
In the latter case  $\chi_T \propto m_q \Sigma$ or $m_{PS}^2 f_{PS}^2$,
so  one ought to see scaling as $\chi_T \propto N_c m_{PS}^2$ at small $N_c m_{PS}^2$.
In fact, there is an old
prediction of a functional form for all mass values, due to Di Vecchia and Veneziano
 \cite{DiVecchia:1980yfw} and Leutwyler and Smilga \cite{Leutwyler:1992yt},
\bee
\chi_T = \frac{m_q \Sigma}{N_f}[\frac{\chi_Q}{\chi_Q + m_q \Sigma/N_f}]
\label{eq:usual}
\ee
or
\bee
\frac{1}{\chi_T} = \frac{N_f}{m_q\Sigma} + \frac{1}{\chi_Q}.
\label{eq:interpolate}
\ee
(The small mass limit of this formula was also derived by Crewther \cite{Crewther:1977ce}.)
With $2 m_q \Sigma = f_{PS}^2 m_{PS}^2$ (appropriate to the $f_{PS}=93$ MeV convention), Eq.~\ref{eq:interpolate} becomes
\bee
\frac{1}{\chi_T} = \frac{2N_f}{f_{PS}^2m_{PS}^2} + \frac{1}{\chi_Q},
\ee
and with $f_{PS}(N_c)= \sqrt{N_c}f_0$ the expected scaling behavior of the pseudoscalar
 decay constant across $N_c$,
 we can write
\bee
\frac{1}{\chi_T} = \frac{2N_f}{N_cf_0^2m_{PS}^2} + \frac{1}{\chi_Q}.
\label{eq:interpolate3}
\ee
That is, the inverse topological susceptibility rises linearly from its
quenched value with respect to the scaling
variable $1/(N_c m_{PS}^2)$ or $1/(N_c m_q)$.

The purpose of this paper is to take a first look at $\chi_T(m_q,N_c)$ -- as the title says, ``a pilot study.''
This means
\begin{itemize}
\item $N_f=2$
\item $N_c=3$, 4, 5
\item One lattice spacing (loosely speaking), roughly matched across $N_c$ using a gluonic observable
(alternatively, roughly matched in bare 't Hooft coupling $\lambda = g^2 N_c$)
\item One simulation volume, a range of intermediate mass fermions, and moderate statistics, so 
all observations are still tentative
\end{itemize}

The goal of the paper is to answer a set of physics questions and a set of simulation questions.
The physics questions are
\begin{enumerate}
\item Does $\chi_T(m_q,N_c)$ actually scale as  $\chi_T(m_qN_c)$ (equivalently $\chi_T(m_{PS}^2 N_c)$),
smoothly connected to $\chi_Q$ at large $m_{PS}^2 N_c$?
\item  Does $\chi_T(m_q,N_c)$ follow the Di Vecchia, Veneziano, Leutwyler, Smilga functional form?
\end{enumerate}
The answers are (1) yes, apparently and (2) qualitatively, but not quantitatively, at the lattice spacings studied.

The main simulation question is: it is well known that in ordinary $N_c=3$ QCD $\chi_T$
has a  very long simulation autocorrelation time $\tau$. How severe an issue is this across $N_c$?
The answer is: $\tau$ grows with $N_c$. $N_c=3$ or 4 seem to be manageable
with the naive approach I took to study the problem, but $N_c=5$ already shows clear issues.

\begin{figure}
\begin{center}
\includegraphics[width=0.6\textwidth,clip]{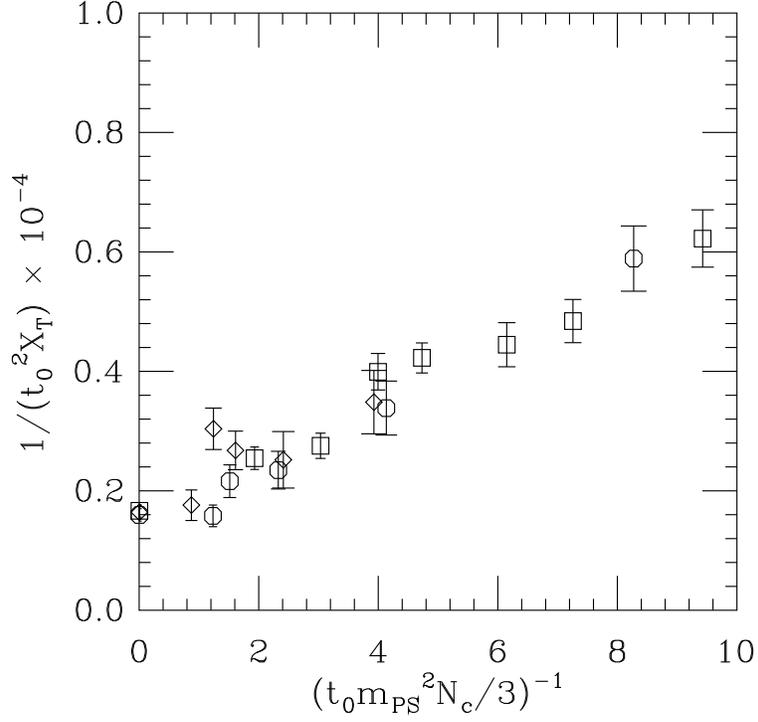}
\end{center}
\caption{The inverse topological susceptibility, scaled by $t_0^2$, versus
$(t_0 m_{PS} ^2N_c/3)^{-1}$. Data are squares for $N_c=3$, octagons for $N_c=4$
and diamonds for $N_c=5$. 
\label{fig:1overtopo}}
\end{figure}

\begin{figure}
\begin{center}
\includegraphics[width=0.6\textwidth,clip]{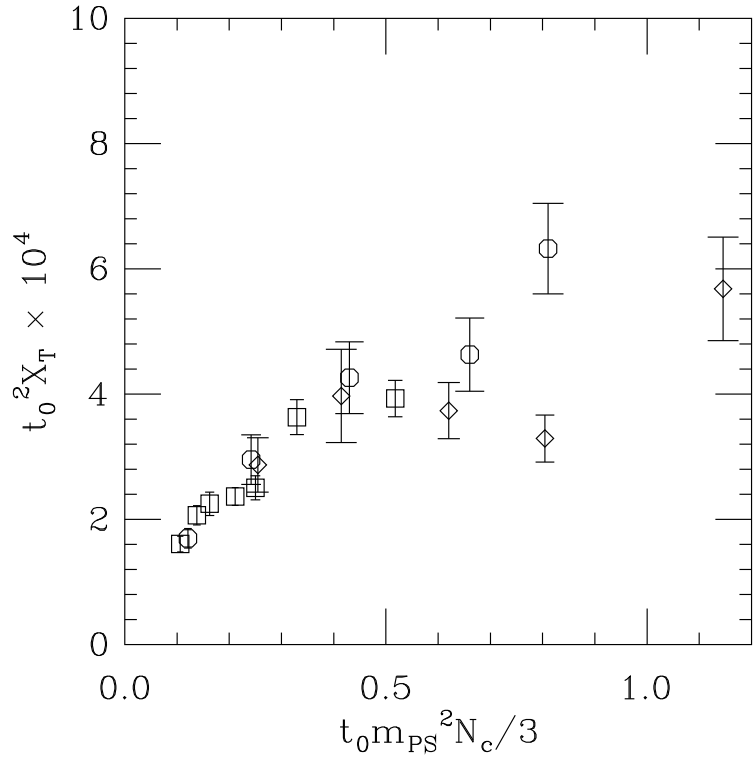}
\end{center}
\caption{A more conventional presentation of the data: $t_0^2 \chi_T$ versus 
$t_0 m_{PS} ^2N_c/3$.
\label{fig:toponcconv}}
\end{figure}

The result of the simulations described here is displayed in Figs.~\ref{fig:1overtopo}
and \ref{fig:toponcconv} (the
 overall scale is set by the ``flow parameter'' $t_0$).
Monte Carlo results collapse to a common curve, which is a straight line in
Fig.~\ref{fig:1overtopo}.
 In that figure we see that the line extrapolates to the
quenched topological susceptibility measured by Ref.~\cite{Ce:2016awn}.
Evidently, the effects of dynamical fermions for this observable  do not depend 
separately on $N_c$ and the fermion mass, but on the combination $N_c m_q$ or $N_c m_{PS}^2$.

In 2001 D\"urr \cite{Durr:2001ty}
presented a similar plot, and comparison to Eq.~\ref{eq:interpolate3}, with $N_c=3$ data.

\begin{figure}
\begin{center}
\includegraphics[width=0.6\textwidth,clip]{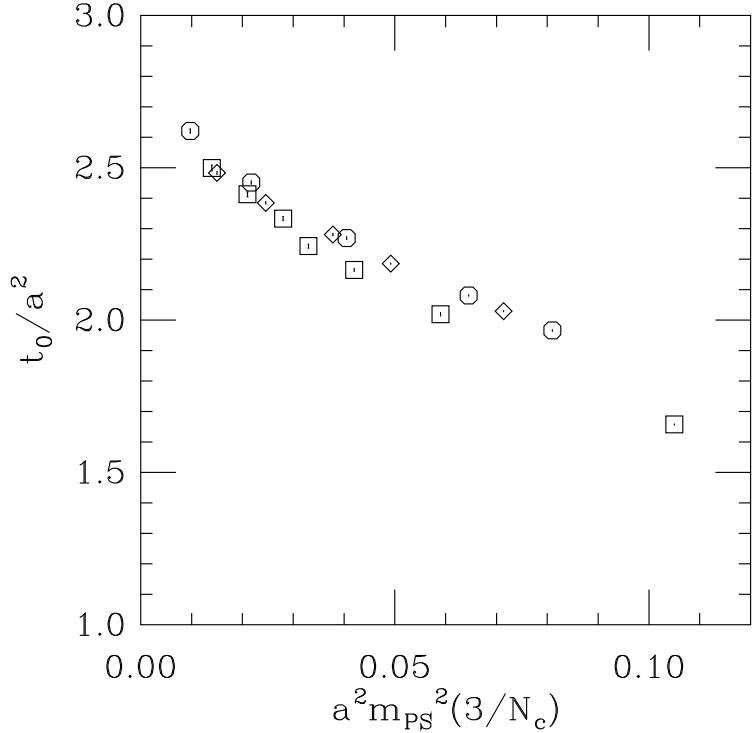}
\end{center}
\caption{The flow scale $t_0(m_{PS})$,
 versus $1/N_c$ times the squared pseudoscalar mass in
lattice units, $(am_{PS})^2/N_c$,
for $N_c=3$ 
 (squares), 4 (octagons), and 5 (diamonds).
\label{fig:t0vsmpi2re}}
\end{figure}

As a contrast, Fig.~\ref{fig:t0vsmpi2re} shows the flow scale $t_0$ versus $(am_{PS})^2/N_c$,
the other non-factorizing  mass and $N_c$ dependence. I am just showing it in passing, since it
has been discussed before.

The outline of the paper is as follows:
Section \ref{sec:tech} describes the calculations: it covers data sets, simulation methodology, and has a discussion of
gradient flow based observables. Here is where I describe how I dealt with the long autocorrelations in the data.
Section \ref{sec:results} presents results: first, comparisons of my data to high statistics 
calculations of the quenched topological susceptibility, then a comparison to previous calculations of the $N_f=2$
$SU(3)$ susceptibility. These are checks to make sure that the present calculation seems to be in order.
I then discuss my results for the  $N_f=2$ susceptibility across $N_c$.
Section \ref{sec:conclusions} is a brief summary. A reminder of the derivation of Eq.~\ref{eq:interpolate} is given in an appendix.

\section{Technical aspects of the calculation \label{sec:tech}}
\subsection{Simulation methodology,  lattice actions, data sets \label{sec:action}}

The dynamical fermion simulations contained two degenerate flavors of 
Wilson--clover fermions. The gauge action is the usual plaquette action,
with the bare gauge coupling $g_0$ parameterized by  $\beta = 2N_c / g_0^2$.
The fermion action uses gauge connections defined as normalized hypercubic (nHYP)
 smeared links~\cite{Hasenfratz:2001hp,Hasenfratz:2007rf,DeGrand:2012qa} (with the
arbitrary $N_c$ implementation of Ref.~\cite{DeGrand:2016pur}).
The bare quark mass $m_0^q$ is introduced via the hopping
parameter $\kappa=(2m_0^q a+8)^{-1}$. The clover coefficient is fixed to
 its tree level value, $c_{\text{SW}}=1$.
The updating scheme is
 the Hybrid Monte Carlo (HMC)  algorithm \cite{Duane:1986iw,Duane:1985hz,Gottlieb:1987mq}
with a multi-level Omelyan integrator \cite{Takaishi:2005tz} and
multiple integration time steps \cite{Urbach:2005ji}
with one level of mass preconditioning for the fermions \cite{Hasenbusch:2001ne}.

 All lattice volumes are $16^3\times 32$ sites. The gauge fields experience periodic
boundary conditions; the fermions are periodic in space and antiperiodic in time.

All data sets are 5000 to 6000 trajectories in length.
Lattices used for analysis are spaced a minimum of 10 HMC time units apart,
so individual bare parameter sets
contain 490-600 stored lattices.
All data sets (individual $(\beta,\kappa)$ values)  are based on a single stream.

 The data sets were collected at approximately 
equal values of lattice spacing. 
(The bare gauge coupling is fixed at each $N_c$ and only $\kappa$ is varied.)
This precludes a discussion of lattice artifacts.
However, comparisons across $N_c$, or with large $N_c$ phenomenology, can be done at any value
 of the lattice spacing.

The data sets are extensions of ones presented in Refs.~\cite{DeGrand:2016pur,DeGrand:2017gbi}
and full spectroscopy is presented in the first of these references.
Table ~\ref{tab:parameters} summarizes relevant information for the runs.
Across the data sets, $m_{PS}^2$ the squared pseudoscalar meson mass is roughly linear
in the Axial Ward Identity fermion mass $m_q$. The ratio $(m_{PS}/m_V)^2$ where $m_V$ is the
vector meson mass, spans the range 0.16-0.64.

\begin{table}
\begin{tabular}{c c c c c}
\hline
$\kappa$  & $am_q$ &  $(a\,m_{PS})^2$ & $t_0/a^2$ & N \\
\hline
$SU(3)$ $\beta=5.4$ & & & &  \\
\hline                                   
0.1250 & 0.105 &  0.312(2)&  1.657(3)  & 500 \\
0.1265 & 0.059 &  0.163(2)&  2.019(6)  & 500 \\   
0.1270 & 0.042 &  0.116(2)&  2.165(6)  & 500 \\
0.1272 & 0.033 &  0.094(2)&  2.243(7)  & 500\\  
0.1274 & 0.028 &  0.070(2)&  2.333(7)  & 500 \\
0.1276 & 0.021 &  0.057(1)&  2.413(8)  & 500 \\
0.1278 & 0.014 &  0.042(1)&  2.500(9)  & 500 \\   
\hline
$SU(4)$ $\beta=10.2$ & & & &  \\
\hline                                                       
0.1245 & 0.108 &  0.309(1)&  1.966(4)  & 490 \\
0.1252 & 0.086 &  0.238(2)&  2.081(3)  & 490 \\
0.1262 & 0.054 &  0.142(1)&  2.269(4)  & 490 \\
0.1270 & 0.029 &  0.074(1)&  2.451(5)  & 500 \\
0.1275 & 0.013 &  0.035(1)&  2.621(7)  & 500 \\
\hline
$SU(5)$ $\beta=16.4$ & & &  \\
\hline
0.1240 & 0.119 &  0.339(1)&  2.029(2)  & 590 \\
0.1252 & 0.082 &  0.221(1)&  2.185(3)  & 590 \\
0.1258 & 0.063 &  0.163(1)&  2.281(4)  & 490 \\
0.1265 & 0.041 &  0.104(1)&  2.385(4)  & 490 \\
0.1270 & 0.025 &  0.062(0)&  2.483(4)  & 490 \\
\hline
 \end{tabular}
\caption{ $N_f=2$ dynamical fermion data plotted in the figures. 
The column labeled by $N$ gives the number
of lattice analyzed for $t_0$ and $\chi_T$.
\label{tab:parameters}}
\end{table}

\subsection{Gradient flow for length scale\label{sec:flow}}
The lattice spacing and the topological charge are measured using the  technique of
gradient flow \cite{other,Luscher:2010iy},
 a smoothing
method for gauge fields via diffusion in a fictitious (fifth dimensional) time $t$.
 In  continuum language, a smooth gauge field $B_{t,\mu}$ is constructed through  an iterative process
\beea
\partial_t B_{t,\mu} &=& D_{t,\mu}B_{t,\mu\nu}  \nonumber \\
B_{t,\mu\nu} &=& \partial_\mu B_{t,\nu} - \partial_\nu B_{t,\mu} + [B_{t,\mu},B_{t,\nu}] , \nonumber \\
\label{eq:flow1}
\eea
beginning with  the original one,
\bee
B_{0,\mu}(x)=A_\mu(x).
\ee
A squared length $t_0$ is defined through the
field strength tensor built using smoothed degrees of freedom,  $G_{t,\mu\nu}$, using the observable
\bee
\svev{E(t)} = \frac{1}{4}\svev{G_{t,\mu\nu}G_{t,\mu\nu}}.
\label{eq:def}
\ee
It is set by fixing the quantity $t_0^2 \svev{E(t_0)}$
to some value $C(N_c)$
\bee
t_0^2 \svev{E(t_0)} = C(N_c).
\label{eq:flow}
\ee
The choice of $C(N_c)$ across $N_c$ is somewhat arbitrary, as it is for $N_c=3$.
There is a natural set of choices motivated by the perturbative expansion
\beea
t^2 \svev{E} &=& \frac{3}{32\pi}(N_c^2-1) \alpha(q)[1 + k_1 \alpha + ...] \nonumber \\
             &=& \frac{3}{32\pi} \frac{N_c^2-1}{N_c} (4\pi \lambda(q)) [1 + k_1 \alpha + ...] \nonumber \\
\label{eq:pertflow}
\eea
where $\alpha(q)$ is the strong coupling constant at momentum
scale $q \propto 1/\sqrt{t}$ and $\lambda(q)$ is the corresponding 't Hooft coupling.
The large $N_c$ limit, where matching gluonic observables is achieved by matching
the bare 't Hooft couplings, is $t^2\svev{E} \propto N_c$. Beyond that, there are
many possible choices.
In Ref.~\cite{DeGrand:2017gbi}, I tested the leading $C(N_c)\propto N_c$ behavior by matching
$t_0$ to another gluonic observable, an inflection point on the static potential called $r_1$.
This choice amounts to fixing the inflection point across $N_c$.
Most other people adopt a different convention,
\bee
 C(N_c)= C(3) \left( \frac{3}{8} \frac{N_c^2-1}{N_c}\right),
\label{eq:ce}
\ee
taking $C(3)=0.3$ as the usual value used in  $SU(3)$. This amounts to saying that the ratio
$t_0/r_1$ has a $1/N_c$ variation away from its $N_c=3$ value (or, the large $N_c$ limit is different from the
$N_c=3$ ratio), nothing more.
I will follow this choice, rather than the one of Ref.~\cite{DeGrand:2017gbi}, because I want to match
the quenched results  of Ref.~\cite{Ce:2016awn}, and they use the convention of Eq.~\ref{eq:ce}.

The extraction of $t_0$ from lattice data is standard and is described in Ref.~\cite{DeGrand:2017gbi}.
 The gradient flow differential equation is integrated numerically  as described by
 L\"uscher \cite{Luscher:2010iy}.
 Calculations used the usual ``clover'' definition of $E(t)$.
An autocorrelation analysis will described after the next subsection.

\subsection{Gradient flow for topological charge  -- definitions\label{sec:gflow}}
The topological charge density is defined as
\bee
q_t(x)= -\frac{1}{32\pi^2} \epsilon_{\mu\nu\rho\sigma} \Tr G_{t,\mu\nu}(x) G_{t,\rho\sigma}(x)
\ee
and is computed using gauge fields at flow time $t$ as
\bee
Q(t)=a^4 \sum_x q_t(x)
\ee
In a system with periodic boundary conditions, the topological susceptibility is simply
\bee
\chi_T = \frac{1}{V}\svev{Q(t)^2}.
\label{eq:chisimple}
\ee

This point actually needs a bit more discussion. 
Eq.~\ref{eq:chisimple} implicitly assumes that $\svev{Q(t)}=0$ when averaged over the
measurements taken in the simulation. The observation  of $\svev{Q(t)} \ne 0$ is an artifact,
indicating that the data has long time autocorrelations. To see if this affects my results,
I will compare $\svev{Q^2}$ to the full correlator
$C(t)=\svev{Q(t)^2} - \svev{Q(t)}^2)$.

A second issue is that,  at any nonzero lattice spacing,
$\chi_T(t)$ depends on $t$.  Taking the continuum limit involves measuring $t_0^2\chi(t)$
at several lattice spacings and taking the $a \rightarrow 0 $ limit (this could be done by plotting the data versus $1/t$ or, re-inserting
the lattice spacing $a$, plotting versus $a^2/t_0$). In principle,
this could be done for any $t$.
The data in this study are all at one lattice spacing, so one has to ask,
are the physics hints given by a study at one value of $a$ sensitive to the choice of operator
(choice of $t$ for $Q(t)$).

\subsection{Data analysis \label{sec:autocorr}}

Both $t_0$ and the topological charge show simulation time autocorrelations.
I attempted to estimate the autocorrelation time through the autocorrelation function
  defined as
\bee
\rho_A(\tau) = \frac{\Gamma_A(\tau)}{\Gamma_A(0)}
\ee
(for a generic observable $A$) where
\bee
\Gamma_A(\tau)= \svev{(A(\tau)-\bar A)(A(0)-\bar A)}.
\ee
The integrated autocorrelation time, up to a window size $W$, is
\bee
\tau_{int}(W) = \frac{1}{2} + \sum_{\tau=1}^W \rho(\tau).
\ee
Unless the total length in time of the data set is much larger than  the
autocorrelation time, it is difficult to estimate an error for $\tau_{int}$.
I analyzed my data sets by breaking them into multiple parts, each part being order
1000 trajectories or 100 saved lattices, computing $\tau$ on each part,
and taking an error from the part-to-part fluctuations.

\begin{figure}
\begin{center}
\includegraphics[width=0.8\textwidth,clip]{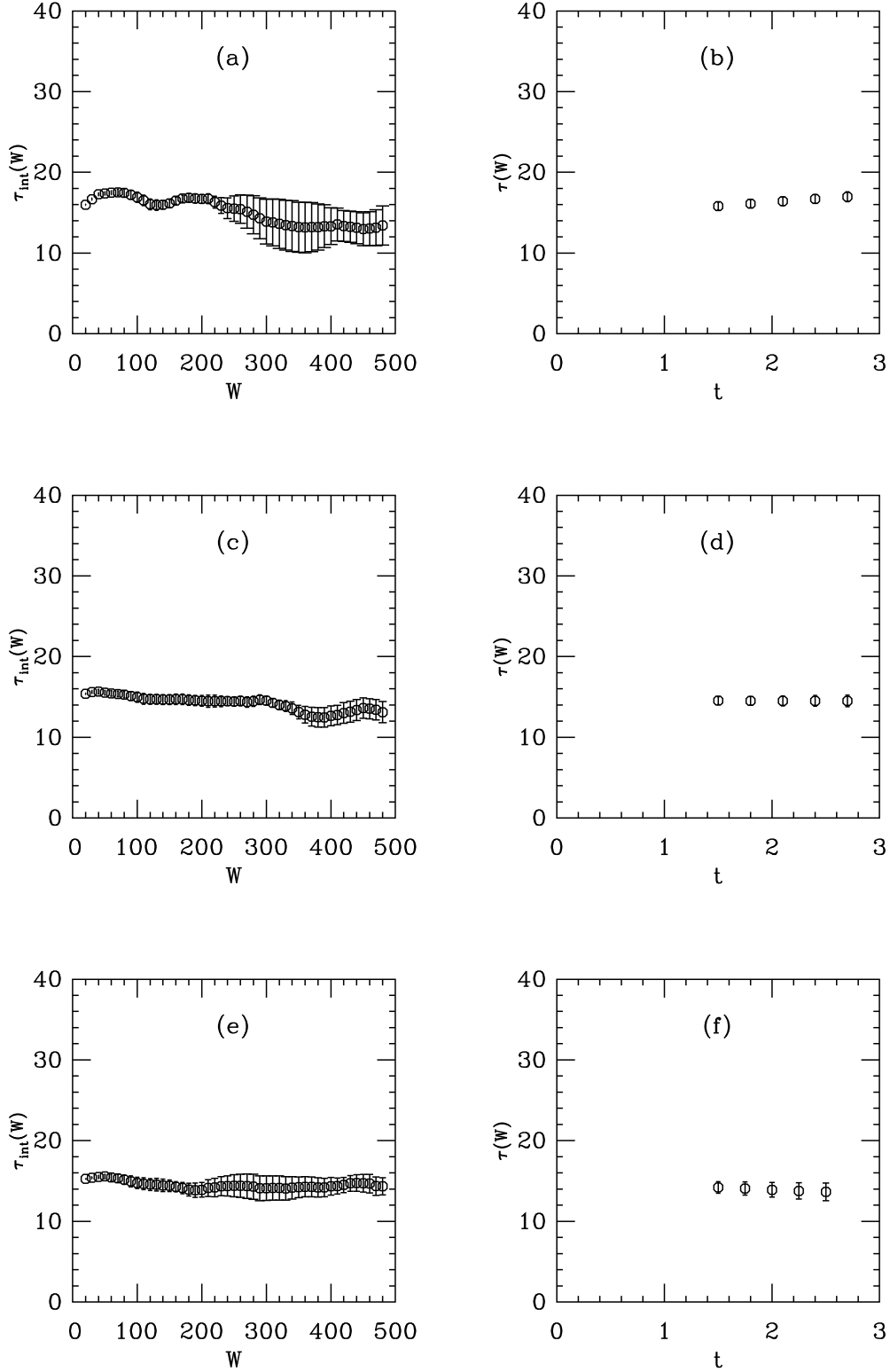}
\end{center}
\caption{ 
Integrated autocorrelation time $\tau_{int}(W)$  for $t^2E(t)$
at fixed $t$ versus $W$ and versus $t$ at fixed $W=200$:
a) and b)  $SU(3)$, $\kappa=0.127$;
c) and d)  $SU(4)$, $\kappa=0.1262$;
e) and f) $SU(5)$, $\kappa=0.127$.
\label{fig:t0autocorr}}
\end{figure}

The analysis of $t_0$ is straightforward. I show a few representative figures,
since the data look quite similar across fermion mass and $N_c$.
Fig.~\ref{fig:t0autocorr} shows  plots of the integrated autocorrelation time
$\tau_{int}(W)$  for $t^2E(t)$
at $t=2.1$ (for $N_c=3$  and 4) and $t=2$ for $N_c=5$ versus $W$, and $\tau_{int}(W=200)$ vs flow time $t$.
The values of $t_0$ in the table are taken from a jackknife analysis dropping two successive lattices,
since these figures indicate that the autocorrelation time is 15-20 trajectories.

\begin{figure}
\begin{center}
\includegraphics[width=0.8\textwidth,clip]{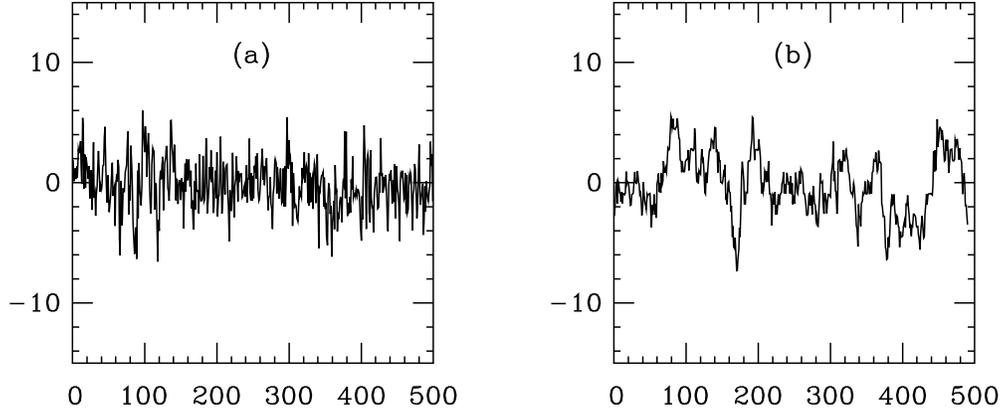}
\end{center}
\caption{ 
Time history of the topological charge at $t=3$ for 
(a) an $SU(3)$ data set ($\kappa=0.1274$)
and
(b) an $SU(5)$ one, $\kappa=0.127$.
\label{fig:topotsu35}}
\end{figure}

Now for the topological charge. The autocorrelation time is large for all $SU(5)$
data sets. This can be seen by eye from 
time histories: compare Figs.~\ref{fig:topotsu35} for an $SU(3)$ history and an $SU(5)$ one.

\begin{figure}
\begin{center}
\includegraphics[width=0.8\textwidth,clip]{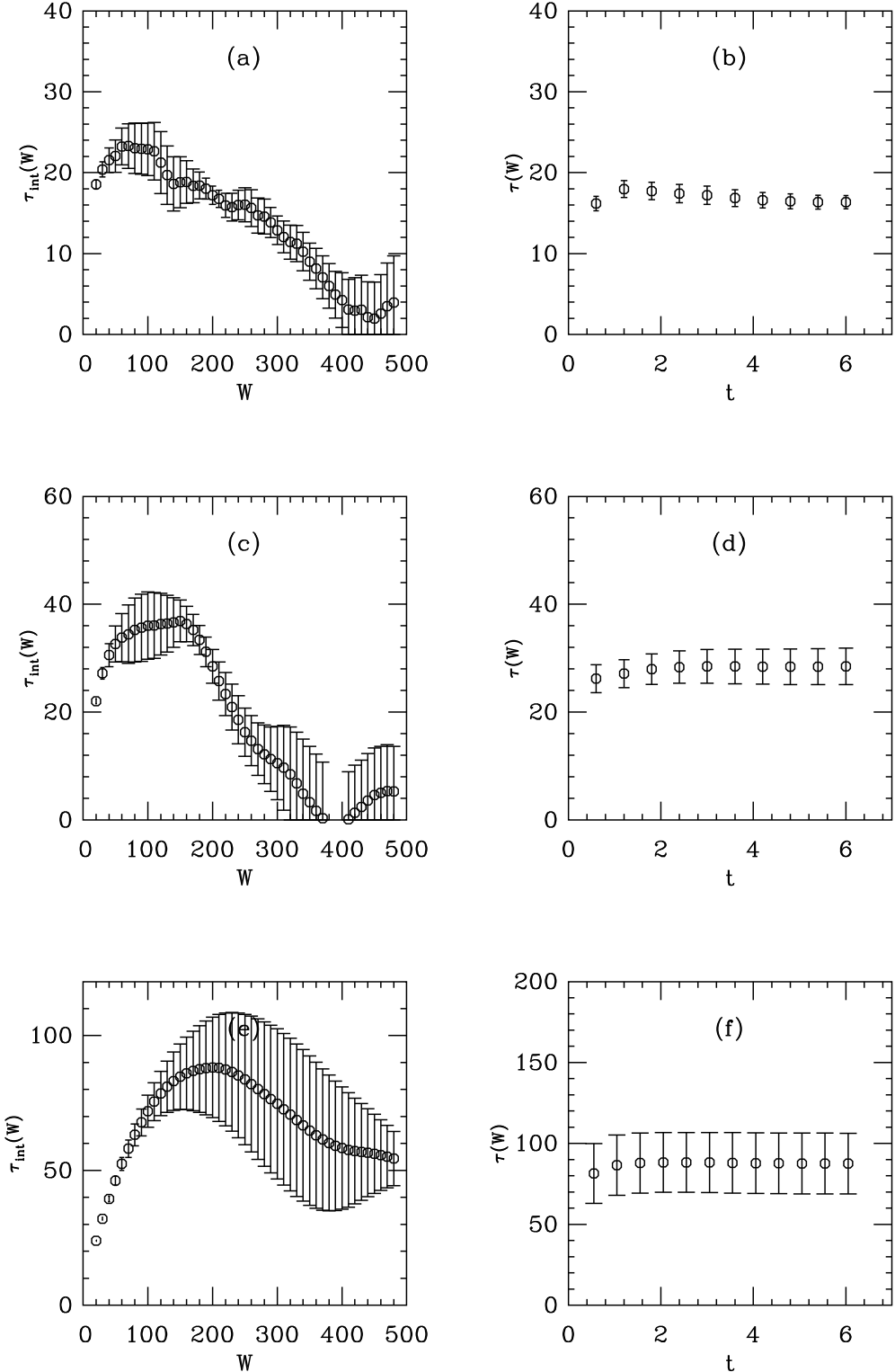}
\end{center}
\caption{ 
Integrated autocorrelation time $\tau_{int}(W)$  for $Q(t)$
at fixed $t$ versus $W$ and versus $t$ at fixed $W=200$:
a) and b)  $SU(3)$, $\kappa=0.127$;
c) and d)  $SU(4)$, $\kappa=0.1262$;
e) and f) $SU(5)$, $\kappa=0.127$.
\label{fig:qautocorr}}
\end{figure}

I repeat the calculation of autocorrelation times for $Q(t)$. In contrast to the results for
$t^2E(t)$, in general $\tau_{int}(W)$ is an irregular function of $W$. This is already an indicator
of long correlations in the data. Results for the same parameter values as in Fig.~\ref{fig:t0autocorr}
are shown in Fig.~\ref{fig:qautocorr}.

\begin{figure}
\begin{center}
\includegraphics[width=0.8\textwidth,clip]{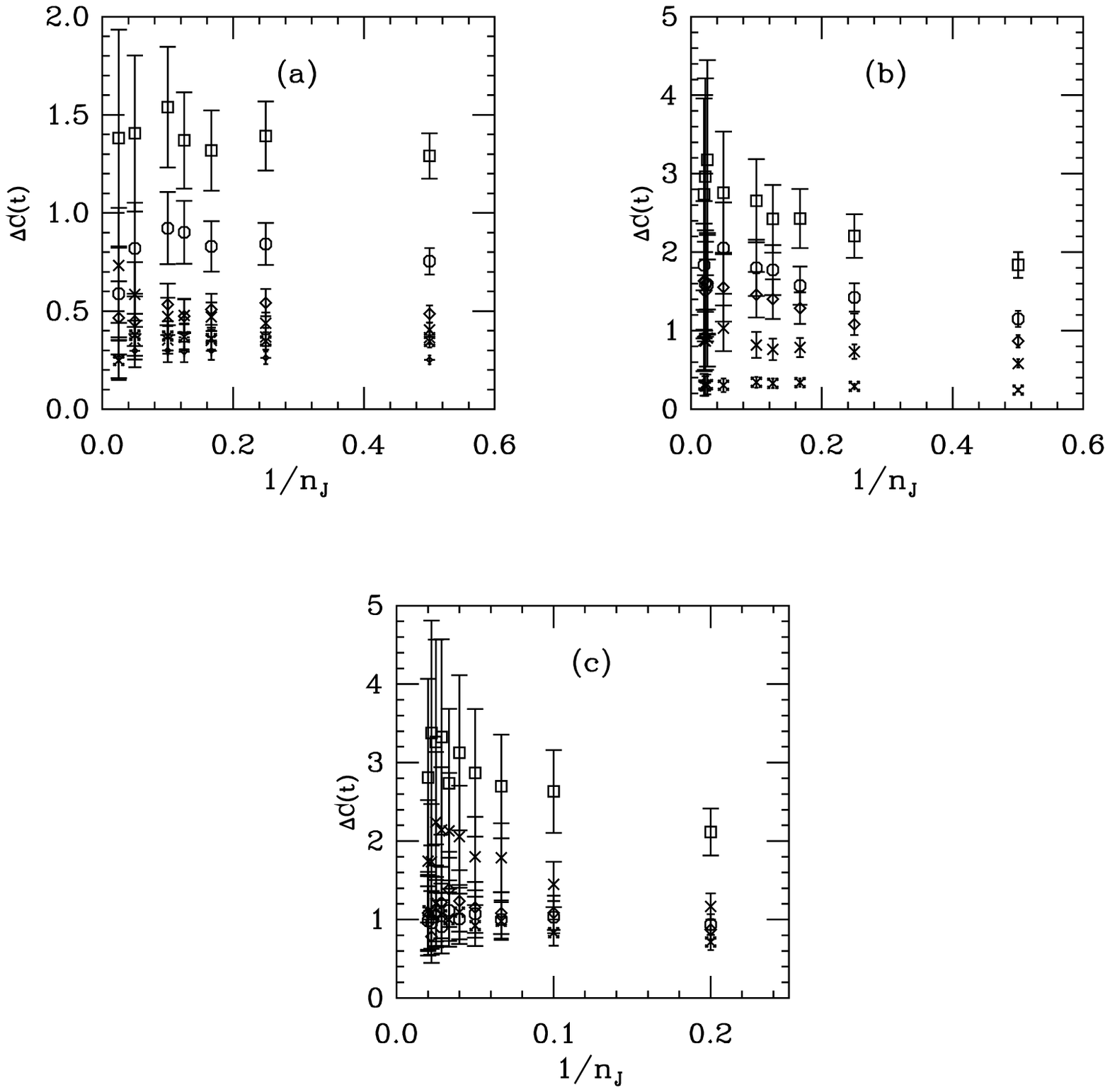}
\end{center}
\caption{ 
Uncertainty in $C(t)$ at $t=3$ as a function of the inverse jackknife size for
(a) $SU(3)$,
(b) $SU(4)$,
and (c) $SU(5)$.
The different plotting symbols correspond to different $\kappa$ values,  the ordering top to bottom is
with decreasing fermion (or pseudoscalar meson) mass.
\label{fig:errorvs1nb}}
\end{figure}

Fit results come from a jackknife analysis, 
removing sets of lattices whose length is longer than the estimated integrated
autocorrelation time.
 This would be $n_J$ successive lattices for $\tau_{int}=10n_J$ molecular dynamics time units.
To be explicit: for a given jackknife I compute the averages $\svev{Q(t)}$, $\svev{Q^2(t)}$ and 
  $C(t) =\svev{Q^2(t)}-\svev{Q(t)}^2$; the uncertainty of each comes from a jackknife.
  I varied the size of the jackknife
beyond the estimate of the integrated autocorrelation time.
I estimate the fractional error from loss of statistics
as
\bee
\frac{\Delta (\Delta C(t))}{\Delta C(t)} = \sqrt{\frac{2}{n}}
\ee
where $n= N/n_J$. That gives a rough error bar. The uncertainty in $C(t)$ increases
with $n_J$ and then either saturates, or at least the growth becomes smaller than
what statistics allows one to see.
Results for $t=3$ are shown are shown in Fig.~\ref{fig:errorvs1nb}. Other $t$ values are similar.
The results of Fig.~\ref{fig:errorvs1nb} suggest the size of the jackknife used to present results.
For $N_c=3$, the autocorrelation analysis suggest an autocorrelation time of about 20 trajectories,
reasonably constant across $\kappa$ values and for $t$ in the range of about 1 to 6, and hence a
 cut $n_J=2$. Fig.~\ref{fig:errorvs1nb} encourages setting the jackknife cut at $n_J=4$.
For $N_c=4$, the autocorrelation time is about 30 trajectories but jackknife errors 
do not saturate until $n_J=8$. Finally, for $N_c=4$, the autocorrelation time is about 50 trajectories
for the two smallest $\kappa$ values and 100 for the others, but Fig.~\ref{fig:errorvs1nb}
instructs us to take $n_J=10$ for the three smallest $\kappa$ values and 20 for the others.

Now for fits to the data. 
I observe, generally, that at small $t$, $Q(t)$ has a Gaussian distribution. At large $t$, individual
configurations ``cool,'' that is, $Q$ peaks at equally spaced, roughly integer values.
This appears to happen at smaller $t$ for $N_c=5$ than it does for $N_c=3$.
I test that the data is Gaussian using the Kolmogorov-Smirnov test \cite{NR}. It compares the integrated
distributions (the cumulants) of the measured data $C(x)$ and the
theoretical prediction $P(x)$.
The cumulant of the measured data is
$C(x) = n(x)/N$ where $n(x)$ is the number of data points with a value smaller than
 $x$ and $N$ the total number of data points.
The theoretical prediction for this quantity is found by integrating
the distribution: $P(x)=\int_{-\infty}^x f(y)  dy$.
The quantity of interest is the largest deviation of $P$ and $C$:
$D=\max_x |P(x)-C(x)|$. From this the confidence level is given by
\bee
Q_{KS}\left((\sqrt{N}D\right)
\label{eq:qks}
\ee
where
\bee
Q_{KS}(x)= 2\sum_{j=1}^\infty (-)^{j-1}\exp(-2j^2 x^2)\, .
\ee
(Note larger $Q_{KS}$ is better.)

\begin{figure}
\begin{center}
\includegraphics[width=0.8\textwidth,clip]{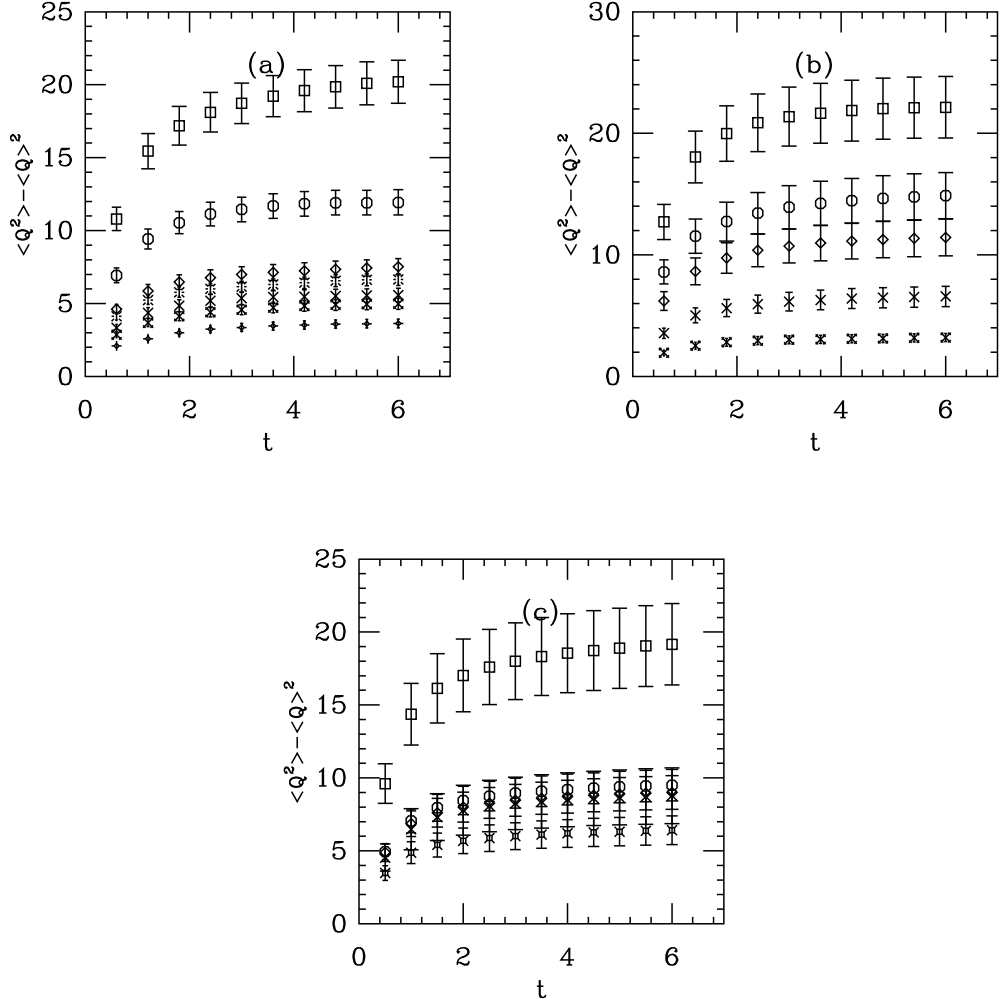}
\end{center}
\caption{ 
$C(t)$ versus $t$ as a function for
(a) $SU(3)$,
(b) $SU(4)$,
and (c) $SU(5)$.
The different plotting symbols correspond to different $\kappa$ values;  the ordering top to bottom is
with decreasing fermion (or pseudoscalar meson) mass.
\label{fig:q2vst}}
\end{figure}

\begin{figure}
\begin{center}
\includegraphics[width=0.8\textwidth,clip]{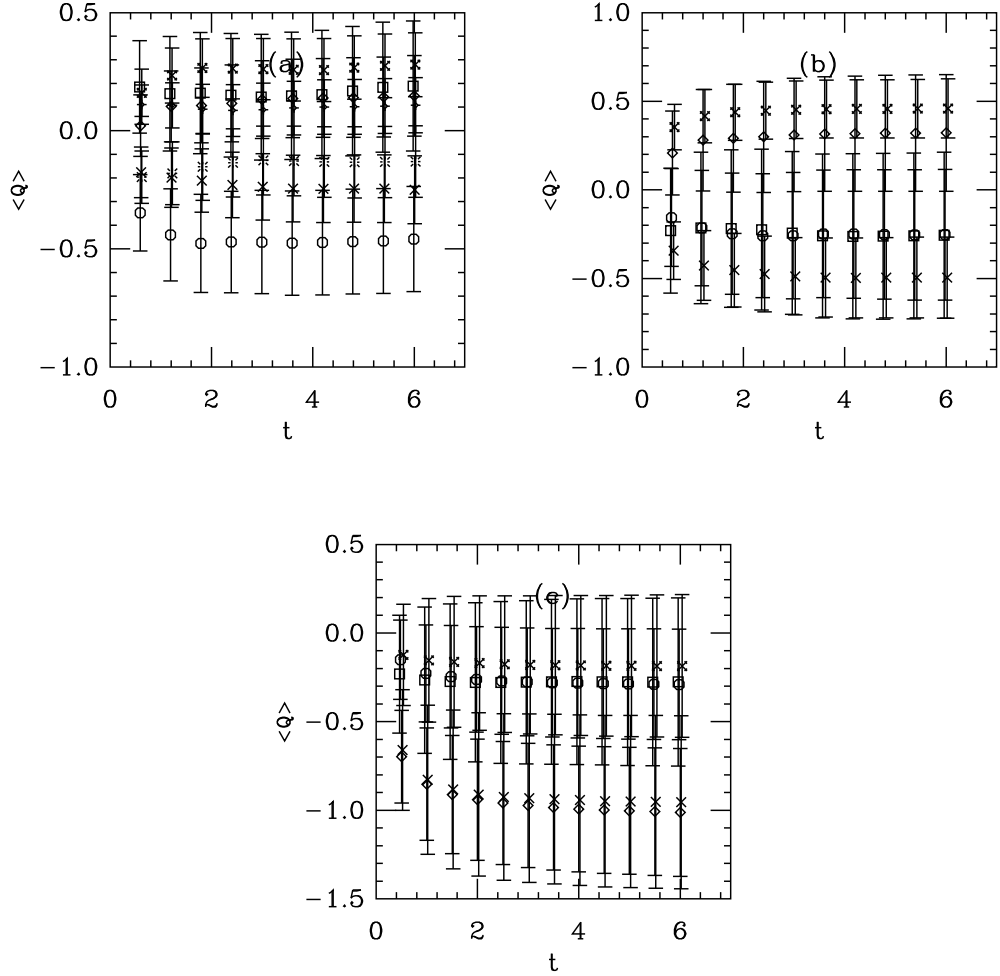}
\end{center}
\caption{ 
$\svev{Q(t)}$ versus $t$ as a function for
(a) $SU(3)$,
(b) $SU(4)$,
and (c) $SU(5)$.
The different plotting symbols correspond to different $\kappa$ values; for numerical values in this cluttered graph,
see Table \protect{\ref{tab:results2}}.
The $x$ axes are slightly displaced for viewing.
\label{fig:qvst}}
\end{figure}

So far, I have not specified a flow time for $C(t)$, so I did fits for a range of $t$ values.
Results for $C(t)$ and $\svev{Q(t)}$ are shown in Figs.~\ref{fig:q2vst} and \ref{fig:qvst}.
Plots of $\svev{Q^2}$ are almost identical to those of $C(t)$.
$\svev{Q(t)}$, in contrast, is nearly independent of $t$.

$Q(t)$ should average to zero. The figures, and the data for $t=3$  presented in Table \ref{tab:results2},
show several cases where $\svev{Q}$ sit two standard deviations away from zero.
However, even for the most extreme deviations ($SU(5)$, $\kappa=0.1265$ and 0.1265) the difference between $C(t)$ and $\svev{Q^2(t)}$
is less than the RMS value of the uncertainties of the two determinations.

In all cases, the data is (nearly) Gaussian about its mean.  This is checked  through a cumulant
analysis where the expectation is (the integral of) a Gaussian with
$\svev{Q}$ and $\svev{Q^2}-\svev{Q}^2$ taken from 
Table \ref{tab:results2}.  Deviations from Gaussianity occur at long flow
time  because $Q$ has cooled to approximate
integers, as revealed by steps in the cumulant.
It didn't seem to be worthwhile to guess a more complicated (Gaussian with steps) distribution for comparison.

\begin{figure}
\begin{center}
\includegraphics[width=0.6\textwidth,clip]{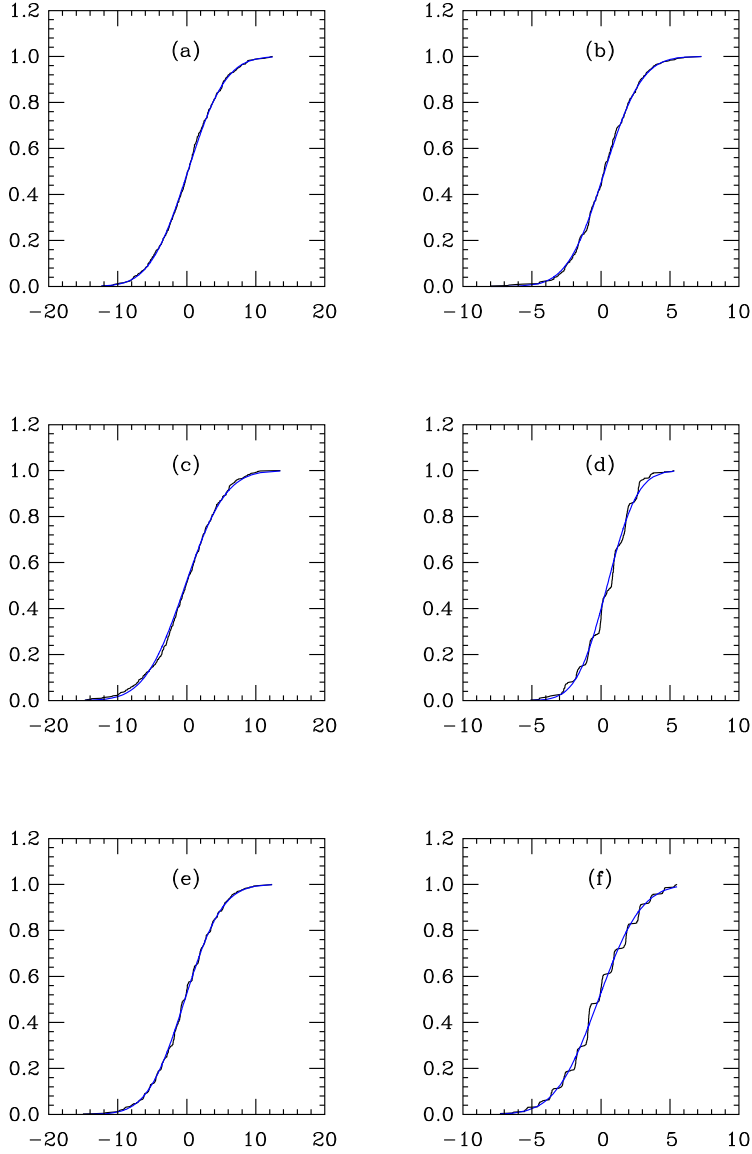}
\end{center}
\caption{ 
Cumulants and fits for selected data sets, all at flow time $t=3$.
 $D$ is the maximum deviation of the cumulant
from the error function and $Q$ is defined in 
Eqs.~\protect{\ref{eq:qks}}. The mean and deviation are taken from 
Table \protect{\ref{tab:results2}}.
$SU(3)$,  (a) $\kappa=0.125$ -- $D=0.027$, $Q=0.84$, (b)  $\kappa=0.1276$ -- $D=0.028$, $Q=0.80$;
$SU(4)$,  (c) $\kappa=0.1245$ -- $D=0.035$, $Q=0.57$,  (d) $SU(4)$, $\kappa=0.1275$ -- $D=0.075$, $Q=0.007$;
$SU(5)$, (e) $\kappa=0.124$ --   $D=0.037$, $Q=0.38$, (f) $\kappa=0.127$ -- $D=0.066$, $Q=0.027$.
\label{fig:cumul}}
\end{figure}

A few pictures of cumulants shown in Fig.~\ref{fig:cumul} illustrate the fits.

Fig.~\ref{fig:q2vst} shows that once $t$ becomes greater than about 2.5, the value of $C(t)$ (and $\svev{Q(t)^2}$, which is almost
identical) roughly forms a plateau. In a better study, I would fix $t$ to any convenient value and extrapolate
$\svev{Q(t)^2}$ to $a=0$. For the remainder of this study I will just fix $t$ to $t=3$.
Results are summarized in Table  \ref{tab:results2}. All phenomenology in the next section
will be done with $\chi_T=\svev{Q(t=3)^2}/V$.

\begin{table}
\begin{tabular}{c c c c}
\hline
$\kappa$   & $\svev{Q}$ & $\svev{Q^2}-\svev{Q}^2$ & $\svev{Q}^2$ \\
\hline
$SU(3)$ $\beta=5.4$ & & &  \\
\hline
0.1250   &  0.14(27) &  18.73(139)&  18.75(139) \\
0.1265   &  -0.47(22) &  11.45(84)&  11.68(90) \\
0.1270   &  0.13(17) &  6.99(54)&  7.00(54) \\
0.1272   &  -0.13(15) &  6.15(37)&  6.17(37) \\  
0.1274   &  -0.24(14) &  5.36(44)&  5.42(45) \\
0.1276   &  0.26(13) &  4.58(35)&  4.65(35) \\
0.1278   &  0.09(11) &  3.36(26)&  3.37(26) \\
\hline
$SU(4)$ $\beta=10.2$ & & &  \\
\hline
0.1245   &  -0.24(46) &  21.38(242)&  21.44(245) \\
0.1252   &  -0.26(36) &  13.92(177)&  14.01(177) \\
0.1262   &  0.31(32) &  10.73(140)&  10.84(146) \\
0.1270   &  -0.49(22) &  6.16(76)&  6.44(86) \\
0.1275   &  0.45(16) &  3.02(33)&  3.24(30) \\
\hline
$SU(5)$ $\beta=16.4$ & & &  \\
\hline
0.1240   &  -0.28(46) &  18.00(263)&  18.09(263) \\  
0.1252   &  -0.28(30) &  8.95(103)&  9.03(103) \\  
0.1258   &  -0.97(35) &  8.46(109)&  9.41(113) \\  
0.1265   &  -0.93(47) &  8.25(180)&  9.15(172) \\  
0.1270   &  -0.18(39) &  6.00(92)&  6.10(93) \\  
\hline
\end{tabular}
\caption{Topological charge and related quantities for $N_f=2$, all at flow time $t=3$.
\label{tab:results2}}
\end{table}

\section{Results \label{sec:results}}

With data sets at one bare gauge coupling per $N_c$ it is hard to quantify lattice artifacts.
I can compare my results to other simulations and ask if they look reasonable.
There are two places where this is done.

\subsection{Comparison with  high precision quenched results \label{sec:qresults}}

The first one is the quenched limit. The authors of Refs.~\cite{Ce:2015qha} and \cite{Ce:2016awn}
published high statistics data for $t_0^2 \chi_T$ for $N_c=2-6$. I collected a data set
much smaller than theirs but comparable to my dynamical sets in size and in lattice spacing, to check against theirs.
It is recorded in Table \ref{tab:qresults}.
My sets are 500 measurements per $N_c$, each spaced 100 sets of sweeps through the lattice, each sweep consisting
of a mix of four 
Brown - Woch  microcanonical over-relaxation steps \cite{Brown:1987rra}
and a Cabibbo - Marinari  heat bath update \cite{Cabibbo:1982zn}, performed
on all $N_c(N_c-1)/2$ $SU(2)$ subgroups of the $SU(N_c)$ link variables.

\begin{figure}
\begin{center}
\includegraphics[width=0.6\textwidth,clip]{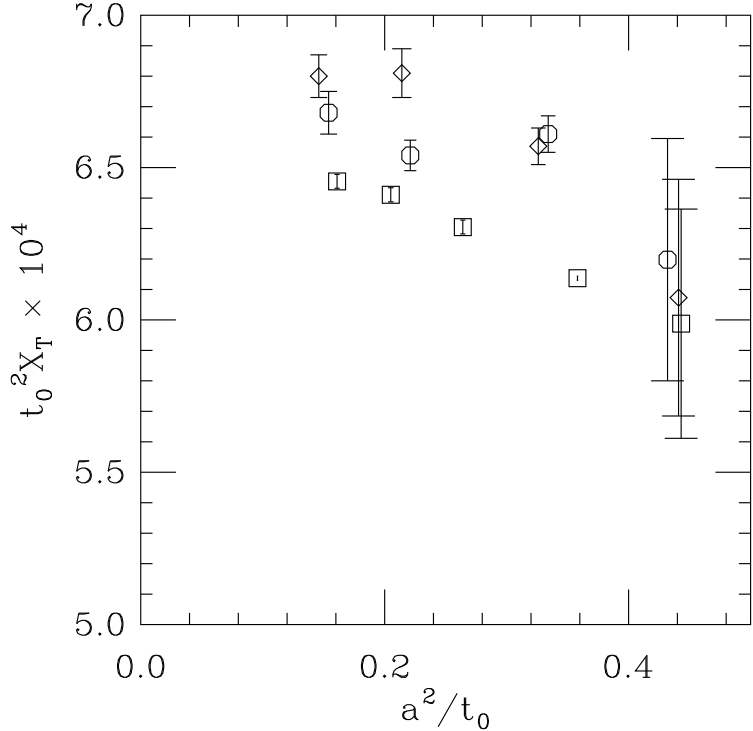}
\end{center}
\caption{Quenched 
$t_0^2 \chi_T$ from Refs.~{\protect\cite{Ce:2015qha,Ce:2016awn}}
(with the small error bars) and by me (with the large error bars), versus $a^2/t_0$.
 Data are squares for $N_c=3$, octagons for $N_c=4$
and diamonds for $N_c=5$.
\label{fig:qscaling}}
\end{figure}

Fig.~\ref{fig:qscaling} shows the comparison. Within my large errors, my results
are compatible with the high statistics results of  Refs.~\cite{Ce:2015qha,Ce:2016awn}.

\begin{table}
\begin{tabular}{c c c c c c}
\hline
$N_c$  & $\beta$  & $t_0/a^2$ & $\svev{Q}$ & $\svev{Q^2}-\svev{Q}^2$ & $\svev{Q}^2$ \\
\hline
 3 & 5.9 &  2.255(10) &  -0.24(17) & 15.43(96) & 15.49(95) \\
\hline
  4 & 10.8 &  2.316(3) &  0.37(17)&  15.17(97) &  15.31(100) \\
\hline
  5 & 17.1 &  2.267(2) &  -0.19(20) & 15.49(99) & 15.52(99) \\
\hline
\end{tabular}
\caption{Quenched data ($\svev{Q}$ etc at $t=3$) plotted in Fig.~\protect{\ref{fig:qscaling}}.
\label{tab:qresults}}
\end{table}

\subsection{Comparison with  high precision $N_f=2$ $SU(3)$ \label{sec:dresults}}

The next comparison is with high precision $N_c=3$, $N_f=2$ results.
I have only been able to find a few recent calculations (most recent studies are
for $N_f>2$ with physical strange (and beyond) fermion masses). But there are three
useful sets.

The first is that of Ref.~\cite{Bruno:2014ova}. I used essentially their techniques:
the topological susceptibility is measured from flow. Ref.~\cite{Bruno:2014ova} presented
data from three small lattice spacings, $a=0.075$ fm, 0.065 fm and 0.048 fm
 (speaking nominally; flow parameters, and hence  the lattice spacing $a$
are computed at each value of bare fermion mass) on very large lattices. The authors of
Ref.~\cite{Bruno:2014ova} provided me with tables of $t_0^2 \chi_T$ versus $t_0 m_{PS}^2$.
Most of their data is at smaller pseudoscalar mass than mine.

\begin{figure}
\begin{center}
\includegraphics[width=0.8\textwidth,clip]{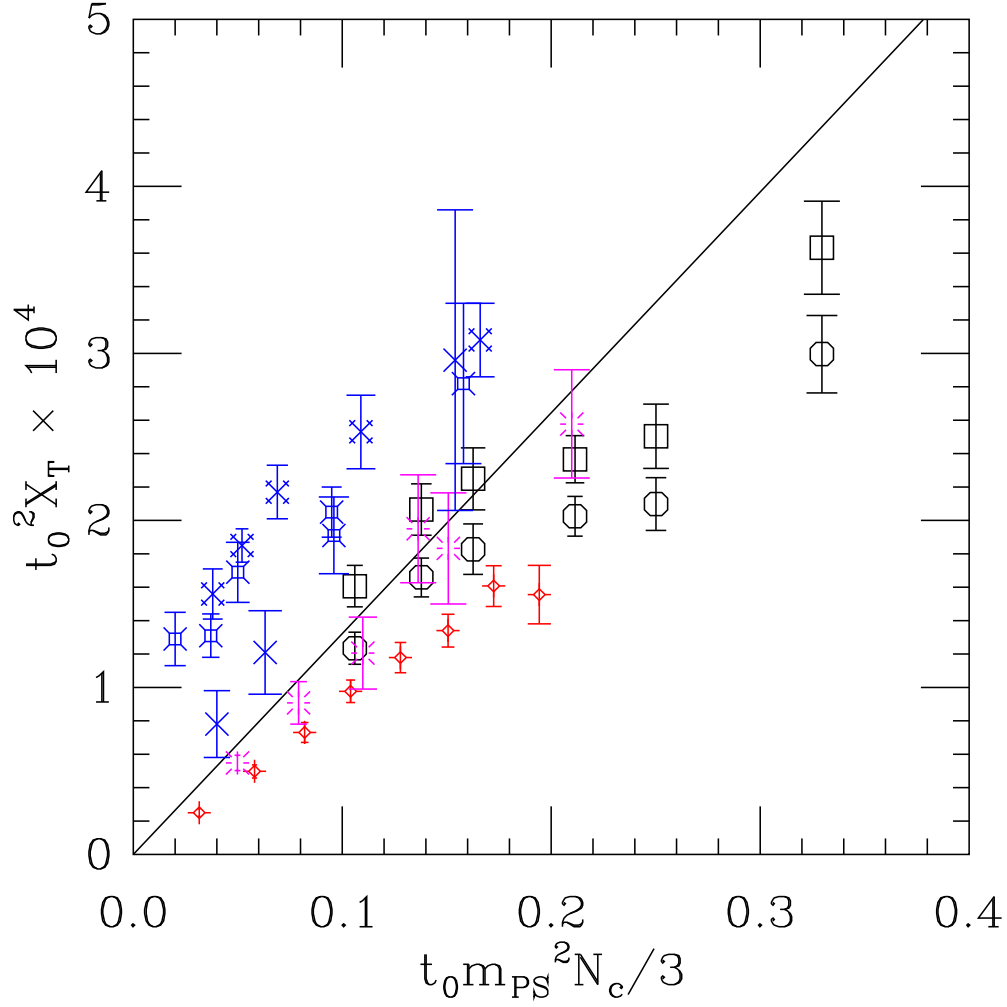}
\end{center}
\caption{Comparison of $N_c=3$ results for $t_0^2 \chi_T$ versus 
$t_0 m_{PS}^2$. My results are black squares for $\chi_T(t=3)$
and black octagons for $\chi_T(t=1.2)$, while the blue points are
data from \protect{\cite{Bruno:2014ova}}: fancy crosses, squares, and crosses
are data at lattice spacing $a=0.075$ fm, 0.065 fm and 0.048 fm, respectively.
Red fancy diamonds are from Ref.~\protect{\cite{Chiu:2011dz}}.
Purple bursts are data from Ref.~\protect{\cite{Aoki:2007pw}}.
The line is $t_0^2 \chi_T =  (t_0 f_{PS}^2/4) m_{PS}^2$ with
$\sqrt{t_0}=0.154$ fm and $f_{PS}=93$ MeV.
\label{fig:compared}}
\end{figure}

The other two calculations measure the topological charge defined using fermionic zero modes.
Ref.~\cite{Aoki:2007pw} is a calculation using overlap fermions in  a sector of fixed
topology.  The lattice spacing is
about 0.12 fm.
They publish a table of $\chi_T r_0^4$ versus $m_{PS}r_0$, where $r_0$ is the
Sommer parameter. \cite{Sommer:1993ce},  an inflection point on the heavy quark potential. 
I take their value $r_0=0.49$ fm and the value of $t_0$ quoted in the review by
Sommer, Ref.~\cite{Sommer:2014mea}, $\sqrt{t_0}=0.154$ fm (from Refs.~\cite{Lottini:2013rfa,Bruno:2013gha})
to rescale the data.
Ref.~\cite{Chiu:2011dz} is a similar calculation with domain wall fermions where the topological
charge is determined using valence overlap fermions. Taking pseudoscalar masses from their
Ref.~\cite{Chiu:2011bm}, I rescale their numbers (quoted in GeV units but determined from $r_0$).
Their data is also shown in Fig.~\ref{fig:compared}. 

The line in the figure is $t_0^2 \chi_T =  (t_0 f_{PS}^2/4) m_{PS}^2$ with
$\sqrt{t_0}=0.154$ fm and $f_{PS}=93$ MeV.
I show my own data for $\chi_T(t)$  for two choices of $t$, 3 and 1.2.

What points am I trying to make with this busy figure? To begin, at the lattice spacings 
of these data sets, lattice artifacts are large and are rather different for the two simulations
based on flow and the ones based on zero modes. The susceptibility measured by flow
is expected to have a lattice artifact $A$ of the form
\bee
t_0^2 \chi_T = b t_0 m_{PS}^2 + A
\label{eq:linearfitchi}
\ee
where $A$ scales as $a^2$. This is what the authors of Ref.~\cite{Bruno:2014ova} saw.
This has been checked in the chiral limit by M\"unster and Wulkenhaar \cite{Munster:2018zdn}.
In contrast, zero modes should drive $\chi_T$ to zero as the fermion mass vanishes.
My own fits to the data of Refs.~\cite{Aoki:2007pw} and ~\cite{Chiu:2011dz} have
intercepts $A=-0.1(1)\times 10^{-4}$ and $-0.04(3)\times 10^{-4}$ respectively,
while the three sets of Ref.~\cite{Bruno:2014ova} are 
$A=1.23(14)\times 10^{-4}$,
$1.02(14)\times 10^{-4}$, and
$0.02(4)\times 10^{-4}$.
Eq.~\ref{eq:linearfitchi} is a good fit to all these data sets.
The interesting quantity in Eq.~\ref{eq:linearfitchi} is $b$, which should
be $b= t_0 f_{PS}^2/4$ from the leading chiral behavior. This is about $13.3\times 10^{-4}$
with $\sqrt{t_0}=0.154$ fm and $f_{PS}=93$ MeV. The line shows this behavior.
A comparison with a ruler shows that the other groups' $SU(3)$ data is consistant with this
value, even though, strictly speaking, $b$ should have its own lattice artifacts
and one would expect agreement only in the continuum limit.

Most of my data is at too large pseudoscalar mass to be expected to be in the
linear regime. At best, the lightest three points might be light enough.
(Note that $m_{PS}L=3.26$ and 3.74 for the two lightest points; smaller $m_{PS}$ would require bigger
simulation volumes than I used, to avoid finite volume contamination.)
This is to be contrasted with the other $SU(3)$ simulations, where $t_0 m_{PS}^2$
is generally lower than 0.16-0.19. My three lowest points lie in the range 0.10-0.16.
Fits to Eq.~\ref{eq:linearfitchi} with more than three points produce $b$ values which
are a factor of two smaller that the expected result, but keeping the lowest
three points produces $(A,b)= 0.36(49)\times 10^{-4}, 12.0(38)\times 10^{-4}   $
 for the $t=3$ susceptibility and
$(A,b)= -0.18(57)\times 10^{-4}   , 13.4(45)\times 10^{-4}   $ for the $t=1.2$ susceptibility.
The two choices should have different lattice artifacts, but the important term ($b$) does not
seem to be a ridiculous value, nor does it seem to be too dependent on the choice of $t$ for $\chi_T$.

\subsection{My results  across $N_c$ \label{sec:myresults}}
Results across $N_c$ were displayed in Figs.~\ref{fig:1overtopo} and \ref{fig:toponcconv}.
Do the data lie on a common curve? I test that by performing a simple linear fit
\bee
\frac{1}{t_0^2\chi} = \frac{1}{t_0^2\chi_Q} + C \frac{1}{t_0(N_c/3)m_{PS}^2}
\label{eq:fitfn}
\ee
 to individual $N_c$ values and to various combinations
of $N_c$. I use my quenched data as inputs to fix the intercept (at $1/(t_0(N_c/3)m_{PS}^2)=0$).
Fit results and the chi-squared per degree of freedom are shown in Table \ref{tab:fitresults}.
The $N_c=3$ and 4 data sets are clearly consistent, and the $N_c=5$ topological susceptibility
falls on the same curve, although the uncertainty in the slope $C$ is clearly much greater.

\begin{figure}
\begin{center}
\includegraphics[width=0.9\textwidth,clip]{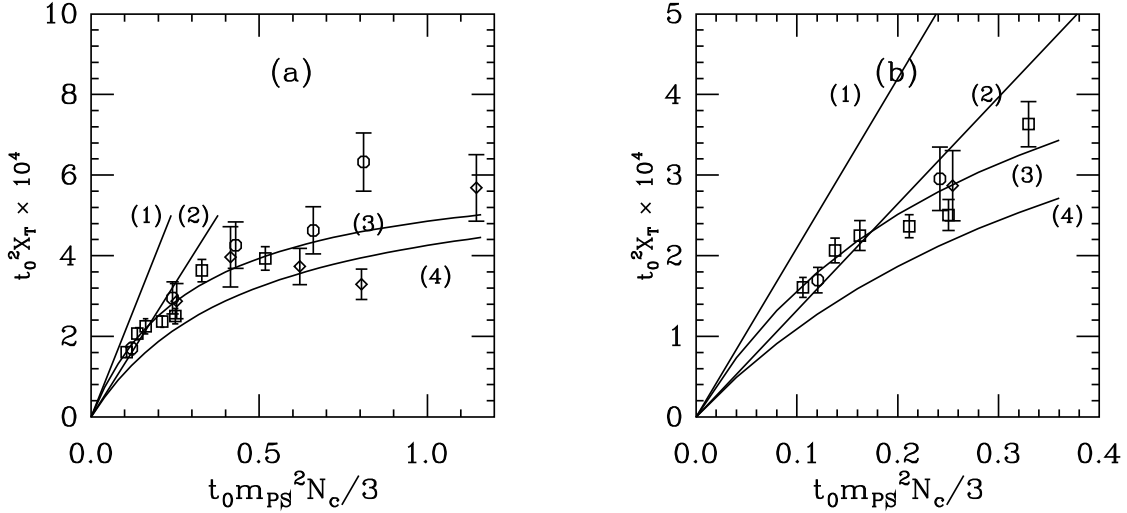}
\end{center}
\caption{ $t_0^2 \chi_T$ versus 
$t_0 m_{PS} ^2N_c/3$ with several lines.
Line (1) is just linear dependence with the slope from the fit to Eq.~\protect{\ref{eq:fitfn}}.
Line (2) is linear dependence with physical values for $t_0$ and $f_{PS}$.
Line (3) is the entire fit function of  Eq.~\protect{\ref{eq:fitfn}}.
Line (4) is the fit function but with physical $C$.
Panel (b) blows up the small mass region from panel (a).
\label{fig:topoline}}
\end{figure}

Fig.~\ref{fig:topoline} replaces the straight-line presentation with a conventional one
of $t_0^2\chi_T$  versus $t_0 m_{PS}^2 N_c/3$.
There are four lines:
Line (1) is just linear dependence with the slope from the fit to Eq.~\protect{\ref{eq:fitfn}}.
Line (2) is linear dependence ($C=4/(t_0 f_{PS}^2)$) with physical  ($SU(3)$) values for $t_0$ and $f_{PS}$.
Line( 3) is the entire fit function of  Eq.~\protect{\ref{eq:fitfn}}.
Line (4) is the fit function but with physical $C$.
Panel (b) blows up the small mass region of panel (a).

\begin{table}
\begin{tabular}{c c c c}
\hline
$N_c$  & $1/(t_0^2\chi_Q)\times 10^4$  & $C\times 10^4$ & $\chi^2$/DoF \\
\hline
3     & 0.165(9) & 0.048(3) & 7.0/6 \\
4     & 0.146(9) & 0.045(5) & 9.4/4 \\
5     & 0.165(10) & 0.051(10) & 7.5/4 \\
3, 4   & 0.155(7) & 0.048(3) & 19.8/11 \\
3, 4, 5 & 0.158(5) & 0.048(3) & 29.1/18 \\
\hline
\end{tabular}
\caption{Results of fits to Eq.~\protect{\ref{eq:fitfn}}. }
\label{tab:fitresults}
\end{table}

\begin{figure}
\begin{center}
\includegraphics[width=0.6\textwidth,clip]{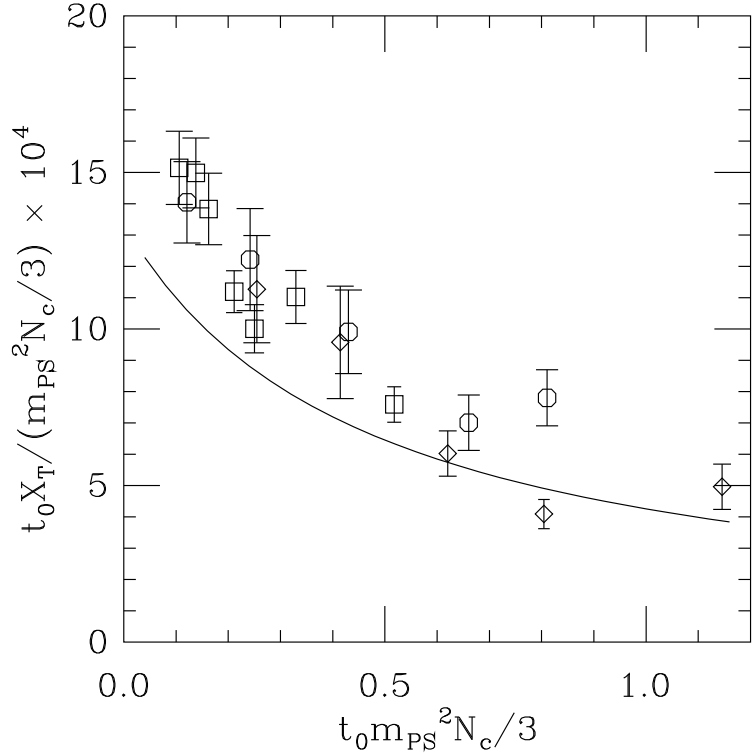}
\end{center}
\caption{ $t_0 \chi_T/(m_{PS} ^2N_c/3))$ versus 
$t_0 m_{PS} ^2N_c/3$ with a line showing the expectation of Eq.~\protect{\ref{eq:interpolate3}}.
\label{fig:chidiv}}
\end{figure}

Finally, Fig.~\ref{fig:chidiv} shows a third view of curve collapse,
 $t_0\chi_T/(m_{PS} ^2N_c/3)$ versus 
$t_0 m_{PS} ^2N_c/3$. This one is a bit dangerous, since $\chi_T$ from flow does
not extrapolate to zero at zero fermion mass: the parameterization blows up there.
Overlaid on the data is the expectation of Eq.~\ref{eq:interpolate3} 
with physical ($SU(3)$) values for $t_0$ and  $f_{PS}$, and $t_0^2\chi_Q$ taken to be a a nominal $6.25\times 10^{-4}$.

What conclusions can be drawn from these figures? First, it's clear that 
$\chi_T$ is, broadly speaking, a function of the combination $m_{PS}^2 N_c$.
Second, it's also clear that Eq.~\ref{eq:interpolate3}
with physical ($SU(3)$) values for $t_0$, $f_{PS}$ and $t_0^2\chi_Q$ taken from
high precision lattice data does not reproduce the data. At this point there
are  two obvious things to say.

First, this difference could just be due to discretization artifacts at the lattice spacing
where the simulations were carried out. A  real check requires several lattice spacings and an extrapolation.

Second, the formula Eq.~\ref{eq:interpolate3} itself could have issues. It is a combination
of lowest order chiral perturbation theory combined with a plausible assumption,
that the eta-prime correlator is a bubble sum. Scaling with $m_{PS}^2 N_c$
is actually scaling with respect to $m_{PS}/\mu_0^2$ where $\mu_0$ is the
eta-prime mass, combined with scaling of $\mu_0^2 \propto 1/N_c$ as expected from the
Witten-Veneziano relation   \cite{Witten:1979vv,Veneziano:1979ec}.  QCD at intermediate to large
sea quark mass does not have to be described by chiral perturbation theory.

\section{Conclusions \label{sec:conclusions}}
This pilot study shows that
fermions influence the topological susceptibility through the product $N_c m_{PS}^2$.
Perhaps it is not a surprising result, but it does illustrate that there are quantities
whose  $N_c$ and fermion mass dependence is non-factorizing.

The other non-factorizing dependence ($\propto m_{PS}^2/N_c$) may be more ubiquitous. It appears in all
chiral logarithm corrections. High quality data for the topological
susceptibility would most likely observe it in the
one loop   \cite{Mao:2009sy}  (and beyond) corrections to $\chi_T$ in the chiral limit.
Probably the easiest place to see this generic behavior is in the dependence of
$t_0$ on the pseudoscalar mass, as shown in Fig.~\ref{fig:t0vsmpi2re}.

Scaling as $N_c m_q$ is expected for observables in the epsilon regime (the limit
of simulation volume $V=L^4$ and pseudoscalar mass
where $m_{PS}L \ll 1$ while $m_H L > 1$ for all other mass scales $m_H$). It appears
in predictions for chiral observables such as the finite-volume condensate
which  involve the scaling combination $m_q \Sigma V$ (for example,   $\Sigma(V)= m_q \Sigma V f(m_q \Sigma V)$). I do not know of any
Monte Carlo checks of this scaling.

This is a pilot study: what would it take to produce higher quality data?
This presumably means larger volumes, several lattice spacings, and maybe larger $N_c$.
Larger volumes are needed to push to smaller fermion mass and check for $m_{PS}^2 N_c$
scaling
in a theoretically clean regime. Several lattice spacings are needed, of course,
 to give a continuum result.
Such data sets already exist for $N_c=3$, and the only reason to repeat them is to use them
as checks of the methodology for the more interesting larger $N_c$ cases.

 I suspect that such $N_c=4$ data sets could be generated
with the same techniques as either I or (better) Ref.~\cite{Bruno:2014ova} used,
simply consuming more resources. (Neglecting autocorrelation effects,
the simulations are dominated by calculation of fermion propagators, involving
matrix-times-vector operations; the scaling is roughly $N_c^2$.)
My experiences with $N_c=5$ raise a flag, however. The long autocorrelation time
for $N_c=5$ compared to lower $N_c$ values is a clear issue. It is hard to imagine the
$N_c>5$ will have a shorter autocorrelation time.
Of course, I should not say more: I have not tried to do extensive running for $N_c>5$ with
dynamical fermions at the same lattice spacing as the data presented here.
But if I were to keep going with this project, I think I would adopt the open boundary
 conditions
used by Ref.~\cite{Bruno:2014ova} to try to shorten the autocorrelation time.

Another ``pilot area'' would be to move away from $N_f=2$. For $N_c$=3, this is reasonably well
explored by simulations with up, down, and strange quarks, and a recent study by
Nogradi and Szikszai \cite{Nogradi:2019iek}
covers $N_f=2-6$. These are all tests at low quark mass: what happens as the mass grows? 
 Varying $N_c$ and $N_f$ together would allow tests of the Veneziano 
limit \cite{Veneziano:1974fa,Veneziano:1976wm},
$N_c \rightarrow \infty$ at fixed $N_f/N_c$.     Is there a universal curve  for $\chi_T(m_{PS}^2)$ 
across a wide range of $N_c$ and $N_f$,  with a scaling variable just $m_{PS}^2N_c/N_f$?


\begin{acknowledgments}
I am grateful to the authors of Ref.~\cite{Bruno:2014ova} for providing me with tables of their data.
Some computations were performed on the University of Colorado cluster.
I would also like to thank Anna Hasenfratz and Oliver Witzel for comments on the manuscript.
My computer code is based on the publicly available package of the
 MILC collaboration~\cite{MILC}. The version I use was originally developed by Y.~Shamir and
 B.~Svetitsky.
This material is based upon work supported by the U.S. Department of Energy, Office of Science, Office of
High Energy Physics under Award Number DE-SC-0010005.
Some of the computations for this work were also carried out with resources provided by the USQCD 
Collaboration, which is funded
by the Office of Science of the U.S.\ Department of Energy 
using the resources of the Fermi National Accelerator Laboratory (Fermilab), a U.S. 
Department of Energy, Office of Science, HEP User Facility. Fermilab is managed by
 Fermi Research Alliance, LLC (FRA), acting under Contract No. DE- AC02-07CH11359.
\end{acknowledgments}

\appendix
\section{Review of the derivation of Eq.~\ref{eq:interpolate} }

Just for completeness, I give a quick derivation of Eq.~\ref{eq:interpolate}.
Since this paper is a lattice calculation, I will
 assume that we have overlap fermions: their zero modes are chiral and 
their nonzero ones are not. With nonchiral lattice fermions, this result will be modified by
lattice artifacts, but let us neglect them for now.
No claim for the originality of this derivation is implied.

\begin{figure}
\begin{center}
\includegraphics[width=0.8\textwidth,clip]{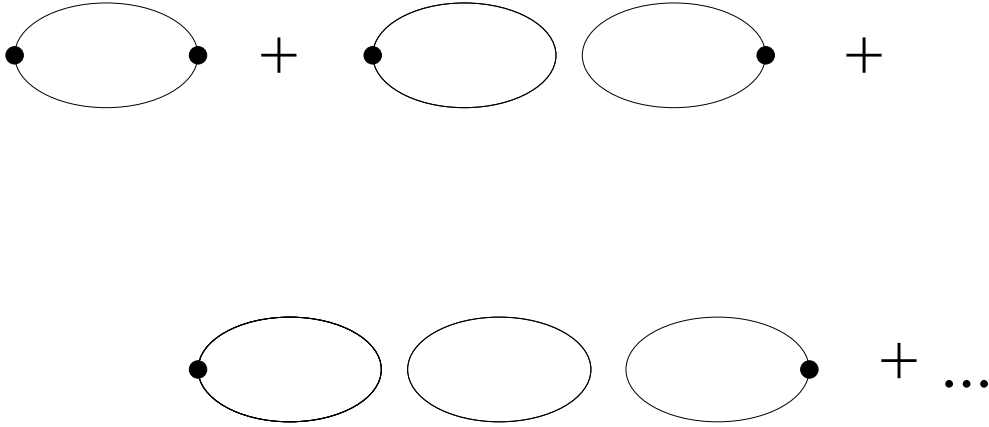}
\end{center}
\caption{
 A set of quark line graphs for the  eta-prime meson.
geometric series to shift the eta-prime mass away from the mass of
the flavor nonsinglet pseudoscalar mesons. The first two terms in the series are the
``connected'' and ``hairpin'' graphs.
\label{fig:hairpin}}
\end{figure}

Consider the propagator for a bound state of a
single flavor of quark, the correlation function of
two local pseudoscalar densities,  $\bar \psi \gamma_5 \psi$.
The first ingredient of the derivation is a plausible assumption for this amplitude, a bubble sum,
shown in Fig.~\ref{fig:hairpin}.
The hairpin diagram, the second term in the sum, is
\bee
H(x,y) = \langle \Tr \gamma_5 \hat D(x,x)^{-1}
\Tr \gamma_5 \hat D(y,y)^{-1} \rangle 
\ee
where $\hat D(x,y)^{-1}$ is the fermion propagator, the inverse of the Dirac operator.
Because only zero modes  of the overlap Dirac operator are chiral, the volume integral of the
hairpin graph is proportional to the zero mode susceptibility
\bee
\frac{1}{ V} \sum_{x,y} H(x,y) =  \frac{ \langle Q^2 \rangle}{V m_q^2 } =
\frac { \chi}{m_q^2 } ,
\label{eqn:HAIR1}
\ee
where $Q$ is just 
 the difference of positive and negative chirality zero modes, $Q=n_+ - n_-$.

In quenched QCD, as described by
quenched chiral perturbation theory,
 there is  an anomalous coupling of two Goldstone bosons
 in the flavor singlet channel,
parametrized by a coupling with the dimensions of a squared mass.
The hairpin graph is analyzed as if each of its quark
loops is a propagator for an ordinary pseudoscalar Goldstone meson.
That is, the momentum space amplitude for the
 connected graph (the first term in Fig.~\ref{fig:hairpin}) is
\bee
C(q)= f_P \frac{1}{q^2 + m_{PS}^2} f_P
\ee
while the hairpin amplitude involving a single flavor is
\bee
H(q) = f_P \frac{1}{q^2 + m_{PS}^2}
 \frac{\mu_0^2}{ N_f} \frac{1}{q^2 + m_{PS}^2} f_P  .
\label{eqn:HAIR2}
\ee
In these expressions,
 $f_P = \langle 0 |\bar\psi \gamma_5 \psi  | PS \rangle
= \sqrt{2}m_{PS}^2f_{PS}/(2m_q)$ from the PCAC relation.
(Here $f_{PS}=93$ MeV.)
The quantity $\mu^2_0$ which couples the fermion loops is the squared mass of the
``quenched approximation eta-prime'' in the chiral limit.
(The factor  $1/N_f$ converts the single-flavor graph into the expectation
of the eta-prime mass in $N_f$-flavor QCD,
 since each closed loop has a multiplicity of $N_f$,
and the wave function (vertex) is scaled by a factor of $1/\sqrt{N_f}$.)
In full QCD the correlator which gives the mass of the isosinglet
meson is the difference $C(t)-N_f H_{full}(t)$, and $H(t)$ is supposed to
represent  the first
term in a geometric series, the rest of the terms in Fig.~\ref{fig:hairpin}. This series
 sums up to
\bee
C(q) - N_f H_{full}(q) =
C(q) - N_f H(q) + \dots = f_P \frac{1}{q^2+m_{PS}^2+ \mu_0^2}f_P,
\label{eqn:MODESUM}
\ee
shifting the squared mass of the pseudoscalar
meson from $m_{PS}^2$ to $m_{PS}^2 + \mu_0^2$.

Computing the quenched susceptibility directly from
Eq. (\ref{eqn:HAIR2}) gives
\bee
\frac{1}{V} \sum_{x,y} H(x,y) =  \frac{f_P^2}{ m_{PS}^4} \frac{\mu_0^2}{ N_f}
 = \frac{\mu_0^2 f_{PS}^2}{2 N_f m_q^2} .
\label{eqn:HAIR3}
\ee
Equating Eqs. (\ref{eqn:HAIR1})
and (\ref{eqn:HAIR3}), we obtain the Witten-Veneziano \cite{Witten:1979vv,Veneziano:1979ec}
relation $\mu_0^2=2N_f\chi/f_{PS}^2$, where $\chi_Q$ is the quenched zero mode
susceptibility.

In full QCD, with dynamical fermions, Eq.~(\ref{eqn:HAIR1}) gives the
quenched topological susceptibility $\chi_Q$. In full QCD,
the hairpin is still saturated by zero modes, but Eq.~\ref{eqn:MODESUM}
(evaluated at $q^2=0$)
says
\bee
\frac{\chi}{m_q^2} = \frac{f_P^2}{N_f}(\frac{1}{m_{PS}^2} - \frac{1}{m_\eta^2}) .
\ee
Substituting for the condensate via $m_{PS}^2 f_{PS}^2 = 2m_q \Sigma$,
recalling $m_\eta^2=\mu_0^2+m_{PS}^2$,
and using the Witten-Veneziano relation to replace $\mu_0^2$ by $\chi_Q$,
we find
\bee
\chi = \frac{m_q \Sigma}{N_f}[\frac{\chi_Q}{\chi_Q + m_q \Sigma/N_f}]
\ee
or
\bee
\frac{1}{\chi} = \frac{N_f}{m_q\Sigma} + \frac{1}{\chi_Q}.
\ee
This interpolates between the small-$m_q$ suppression 
and the quenched result.

\end{document}